\documentclass[sigconf]{acmart}
\AtBeginDocument{%
  }

\setcopyright{acmlicensed}
\copyrightyear{YYYY}
\acmYear{YYYY}
\acmDOI{XXXXXXX.XXXXXXX}
\acmConference[Conference acronym 'XX]{Make sure to enter the correct
  conference title from your rights confirmation email}{Month DD--DD,
  YYYY}{City, Province}
\acmISBN{978-1-4503-XXXX-X/YYYY/MM}

\usepackage{soul}
\usepackage{enumitem}
\usepackage{array}
\usepackage{mdframed}
\usepackage{booktabs}
\usepackage{graphicx}   % for \resizebox
\usepackage{placeins}   % for \FloatBarrier (optional)
\usepackage{caption}
\usepackage{capt-of}   % <— gives \captionof without changing ACM styles
\begin{document}

\title{LLM-Assisted AHP for Explainable Cyber Range Evaluation} 

\author{Vyron Kampourakis}
\authornote{Corresponing author}
\email{vyron.kampourakis@ntnu.no}
\orcid{0000-0003-4492-5104}
\affiliation{%
  \institution{Norwegian University of Science and Technology}
  \city{2802 Gjøvik}
  \country{Norway}
}

\author{Georgios Kavallieratos}
\orcid{0000-0003-1278-1943}
\email{georgios.kavallieratos@ntnu.no}
\affiliation{%
  \institution{Norwegian University of Science and Technology}
  \city{2802 Gjøvik}
  \country{Norway}
}

\author{Georgios Spathoulas}
\orcid{0000-0003-2947-486X}
\email{georgios.spathoulas@ntnu.no}
\affiliation{%
  \institution{Norwegian University of Science and Technology}
  \city{2802 Gjøvik}
  \country{Norway}
}

\author{Vasileios Gkioulos}
\orcid{0000-0001-7304-3835}
\email{vasileios.gkioulos@ntnu.no}
\affiliation{%
  \institution{Norwegian University of Science and Technology}
  \city{2802 Gjøvik}
  \country{Norway}
}

\author{Sokratis Katsikas}
\email{sokratis.katsikas@ntnu.no}
\orcid{0000-0003-2966-9683}
\affiliation{%
  \institution{Norwegian University of Science and Technology}
  \city{2802 Gjøvik}
  \country{Norway}
}

\renewcommand{\shortauthors}{Kampourakis et al.}

\begin{abstract}
Cyber Ranges (CRs) have emerged as prominent platforms for cybersecurity training and education, especially for Critical Infrastructure (CI) sectors that face rising cyber threats. One way to address these threats is through hands-on exercises that bridge IT and OT domains to improve defensive readiness. However, consistently evaluating whether a CR platform is suitable and effective remains a challenge. This paper proposes an evaluation framework for CRs, emphasizing mission-critical settings by using a multi-criteria decision-making approach. We define a set of evaluation criteria that capture technical fidelity, training and assessment capabilities, scalability, usability, and other relevant factors. To weight and aggregate these criteria, we employ the Analytic Hierarchy Process (AHP), supported by a simulated panel of multidisciplinary experts implemented through a Large Language Model (LLM). This LLM-assisted expert reasoning enables consistent and reproducible pairwise comparisons across criteria without requiring direct expert convening. The framework's output equals quantitative scores that facilitate objective comparison of CR platforms and highlight areas for improvement. Overall, this work lays the foundation for a standardized and explainable evaluation methodology to guide both providers and end-users of CRs.
\end{abstract}





\maketitle

\section{Introduction}
\label{S:intro}

Critical infrastructures (CI) such as power and manufacturing plants, water treatment facilities, and transportation networks are increasingly being targeted by sophisticated cyberattacks~\cite{kamp-cracks-2025, makr2021}. Unlike typical Information Technology (IT) system breaches, successful attacks on Industrial Control Systems (ICS) can cause not only data loss, but also physical damage and cascading disruptions to essential services. For example, the Stuxnet worm~\cite{falliere2011w32}, one of the earliest known cyber-physical attacks, demonstrated the feasibility of degrading centrifuge operations in Iran’s Natanz facility, while the TRITON malware~\cite{di2018triton} targeted safety instrumented systems in a Saudi petrochemical plant. These incidents underscore that ensuring the cybersecurity of CI requires specialized training.

In this context, Cyber Ranges (CRs) have gained prominence as a means of enhancing cyber defense capabilities in realistic, yet controlled, environments. Such environments enable hands-on, performance-based learning and team exercises, provide real-time feedback, and allow participants to practice by responding to cyber incidents in a safe and controlled setting~\cite{Kampourakis2025, vkamp2023}. Essentially, they offer \textit{train-as-you-fight} opportunities where defenders can hone their skills against simulated cyber attacks without risking operational availability~\cite{Glas:TAYF}. For CI, CRs can be tailored to include Supervisory Control and Data Acquisition (SCADA) systems, Programmable Logic Controllers (PLCs), and physical process simulations, covering the entire or part of the operational stack of a typical ICS, so that defenders and operators gain cross-layer knowledge of both cyber IT and Operational Technology (OT) environments~\cite{kavallieratos2019towards}. However, a key question arises: How do we assess the effectiveness and suitability of a given CR?

Apparently, not all CRs are equal in their capabilities. CI-focused scenarios demand high realism, e.g., accurate physics of processes, realistic network traffic including industrial protocols, and may even require Hardware-in-the-Loop (HitL) for fidelity. Moreover, organizations need to ensure that the CR they invest in can support their specific use cases, including training of the workforce, incident response exercises, vulnerability testing, or research~\cite{KKamp2025}. Simply put, a systematic framework to evaluate and compare CR platforms is needed to guide stakeholders in designing, developing, or improving CRs, including those aimed at CI security.

Existing efforts to guide CR selection provide partial solutions. The National Institute of Standards and Technology (NIST) released a CR guide with a checklist of features and considerations~\cite{Nist:guide} to aid organizations in assessing CR offerings. The European Cyber Security Organisation (ECSO) in 2024 expanded on this by publishing a detailed CR features checklist~\cite{ECSO:check}, refining NIST’s criteria. However, both these resources solely enumerate important capabilities and features, without providing a quantitative evaluation framework to measure how well a CR should perform across multiple dimensions or to compare platforms objectively. Overall, there is a lack of standardized methods to assess the overall suitability of a CR to provide realistic and impactful training, both for CI and pure Information and Communication Technology (ICT) scenarios.

\textbf{\textit{Contribution:}} To bridge this gap, we propose a structured evaluation framework for CRs, grounded in Multi-Criteria Decision Analysis (MCDA). The framework identifies a comprehensive set of evaluation criteria for CRs and applies the Analytic Hierarchy Process (AHP) to derive their relative weights and compute an aggregate conformance score. A key methodological feature of this work lies in the integration of an LLM-based simulated multidisciplinary panel to perform the pairwise comparisons required by AHP, drawing on codified knowledge from cybersecurity, ICS/SCADA engineering, and critical infrastructure training roles. This approach ensures consistent, explainable, and reproducible judgments on the relative importance of each criterion. The outcome is a score that can inform decisions, such as selecting between different range platforms or highlighting areas where existing ranges need improvement to better serve their objectives.

The remainder of this paper is organized as follows. The next section reviews the most relevant literature on CR evaluation and related multi-criteria assessment approaches. Section~\ref{sec:framework} presents the proposed evaluation framework, including the defined criteria, the AHP-based weighting methodology, and the integration of LLM-simulated expert reasoning. Section~\ref{S:res} reports the results of the AHP-based weighting process, including the simulated expert comparisons, computed criterion weights, and consistency analysis. Section~\ref{S:app} showcases the applicability of the framework through two representative use cases of CI-oriented CRs. The last section concludes the paper and outlines potential directions for future research.

\section{Related Work}
\label{sec:Eval:apps}

Organizations such as NIST and ECSO guide stakeholders to navigate the CR landscape. NIST’s checklist recommends assessing features like supported use cases, deployment model, scenario customization, standards alignment, scalability, and vendor support~\cite{Nist:guide}. Similarly, ECSO’s checklist~\cite{ECSO:check} refines NIST’s criteria by adding factors relevant to procurement, including cost, scenario library size, sector-specific expertise, and maintenance commitments. Both NIST and ECSO stress that organizations should first define their own requirements before applying these checklists. However, while exceedingly useful, the checklists lack a formal evaluation framework; they indicate whether a feature is present but not its relative importance or performance level. In this direction, an MCDA-based framework could address this gap by quantifying performance across criteria and aggregating results into an overall assessment.

Moreover, it is worth noting the (absence of) connection to established security frameworks. Prominently, the NIST Cybersecurity Framework (CSF) for critical infrastructure does not explicitly cover CRs, but emphasizes the importance of training and simulations under the \textit{Protect} and \textit{Respond} functions~\cite{bianchi2024}. Accordingly, CRs can be seen as a means to fulfill certain CSF subcategories of \textit{PR.AT: Awareness and Training}~\cite{PR:AT}. From this viewpoint, this study's framework could be viewed as a complementary piece that ensures the tools used for such training platforms, e.g., a CR, are up to the task.

Regarding academic literature, until now, only a handful of works have focused on evaluating CRs themselves, as opposed to using CRs to evaluate diverse educational and training scenarios. One set of related work has been evaluating the outcomes of CR exercises, particularly training effectiveness. For example, the work in~\cite{Andreolini2020} proposed a framework to evaluate the performance of trainees in CR exercises (CRX). Their approach involved instrumenting the range with a distributed monitoring architecture to capture the actions of trainees and then modeling these actions as directed graphs. However, it focuses on evaluating the human aspect and indirectly the training content rather than the capabilities of the CR platform.

Another pertinent study was presented in~\cite{Glas:TAYF}, highlighting that systematic evaluation has been largely neglected in CRX research and proposing a taxonomy-based evaluation framework. To demonstrate the applicability of the framework, the authors applied it in a user study with 50 students, evaluating a particular training exercise and identifying its strengths and weaknesses. Altogether, this study provided a structured checklist of criteria similar to a usability evaluation, but for training authenticity. Their taxonomy is aimed at evaluating the quality of the training experience from a learning perspective, whereas we aim to evaluate the CR platform itself, especially its suitability for CI scenarios.

Regarding MCDA techniques, they have been applied in various cybersecurity decision problems, inspiring our approach. Notably, the authors in~\cite{Kampourakis2025} introduced a high-level MCDA-based evaluation formula specifically for CRs, defining a set of criteria, modeling interdependencies among these criteria via a weighted dependency graph, and computing an overall conformance score. However, they acknowledge major limitations regarding the subjectivity of weights, the limited criterion set, and contextual dependence, positioning their work as a stepping stone toward more rigorous frameworks. In a different context, the work in~\cite{Guo2023} applied an AHP-based method for network security situation assessment in Industrial Internet of Things (IIoT) environments, determining weights for security impact factors based on the IIoT system properties using AHP and combining them with data analytics to quantify the overall security posture. The success of AHP in that context demonstrated its suitability for problems where multiple factors must be balanced and expert judgment is necessary to weigh them.

\section{Proposed Evaluation Framework}
\label{sec:framework}

Our framework provides a structured process to evaluate the capability and suitability of a CR for both CI and ICT applications; however, later on sections~\ref{sec:criteria-definition} and~\ref{SS:AHP} we tailor the proposed framework for its application in CI scenarios. The framework is built on two main components: i) a defined set of evaluation criteria, and ii) an AHP-based methodology for weighting these criteria. In this section, we describe the evaluation criteria and their selection rationale, then detail the AHP application, including the construction of the pairwise comparison matrix and the scoring aggregation.

\begin{figure}[t]
\centering
\includegraphics[width=0.6\linewidth]{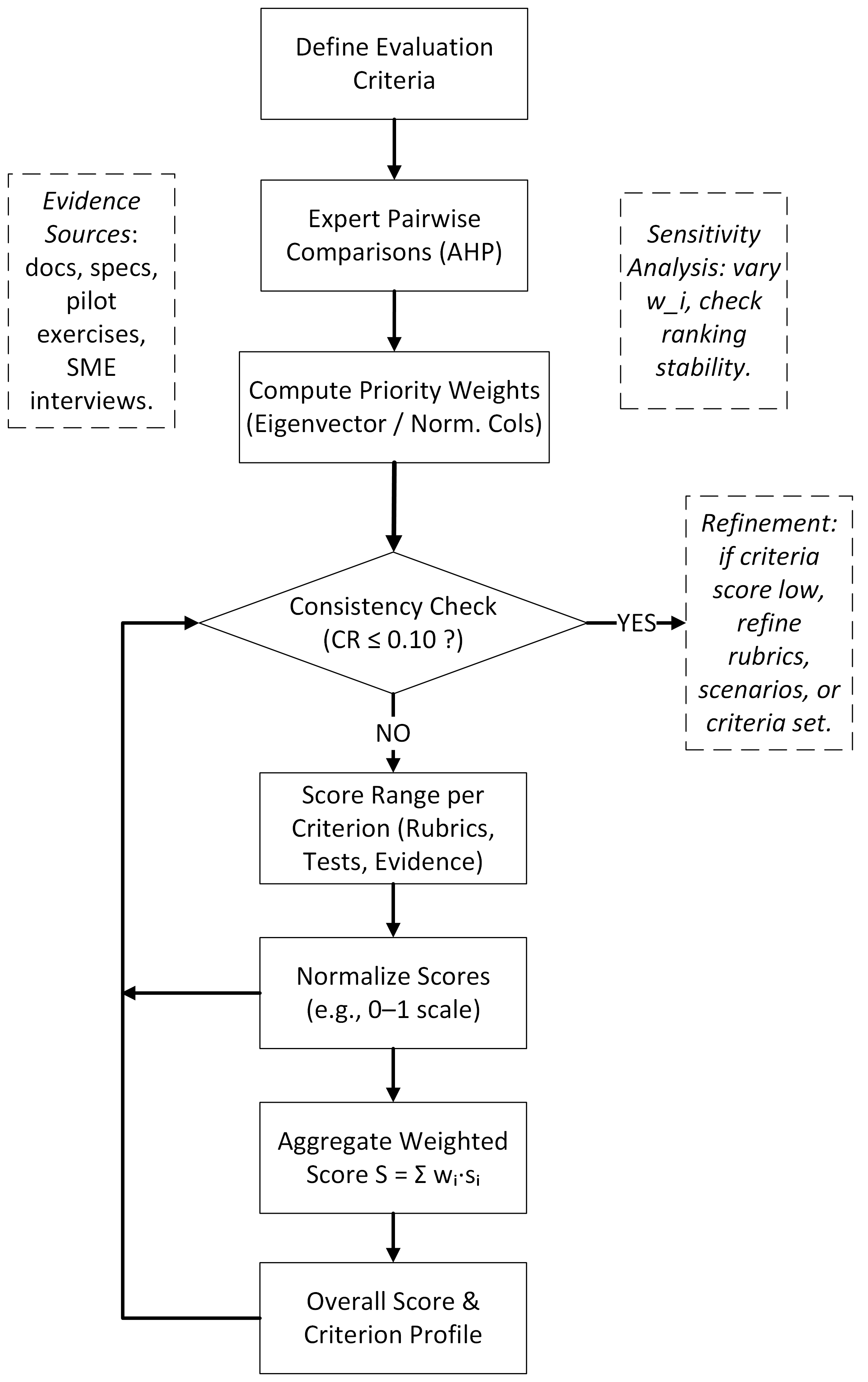}
\caption{Evaluation framework process for CRs.}
\Description{Evaluation framework process for CRs.}
\label{fig:framework}
\end{figure}

Figure~\ref{fig:framework} illustrates the overall architecture of the stepwise framework and its usage workflow. The first step regards the definition of evaluation criteria, establishing the dimensions in which a CR is assessed. These criteria are then subjected to expert pairwise comparisons using the AHP. This step is informed by evidence such as technical documentation, system specifications, pilot exercises, and subject matter input. From these comparisons, priority weights are calculated via eigenvector or normalized column methods, followed by a consistency check.

Once validated weights are obtained, the evaluation continues with scoring the CR on each criterion using predefined anchor rubrics. For consistency reasons, scores are normalized to a common scale (e.g., 0–1) and aggregated into a single overall score computed as the weighted sum of sub-scores across criteria. The final output consists of both a composite score and a detailed criterion profile, allowing evaluators to identify strengths and weaknesses and to adjust them, depending on the case. Side activities embedded in the framework, portrayed as dotted rectangles in figure~\ref{fig:framework}, include sensitivity analysis, where weights are varied to test the soundness of rankings, and iterative refinement, which encourages the adjustment of rubrics, scenarios, or criteria if persistent weaknesses are identified.

\subsection{Evaluation Criteria Definition}
\label{sec:criteria-definition}

In this section, we identify ten overarching evaluation criteria, adapting them for CI scenarios by defining context-specific key indicators and rubric anchors. These criteria synthesize insights from general CR feature checklists~\cite{Nist:guide,ECSO:check} and generic taxonomy studies~\cite{fi17060259, Kampourakis2025, app11041809, YAMIN2020}, combined with CI-specific requirements~\cite{SP-800-82}, and ICS testbed design literature~\cite{ani2019design}. The set is designed to be comprehensive but non-overlapping, so that each criterion captures distinct aspects of the range, making them directly usable with AHP for weighting and with rubric-based scoring for aggregation. Table~\ref{tab:criteria} lists the ten criteria with concise descriptions, CI-specific justification for each criterion, observable indicators suitable for scoring, and rubric anchors to support consistent evaluation. Keep in mind that this list is not exhaustive and can be adapted or extended to fit the specific goals and operational context of a given CR.

\begin{table*}[p] % float page for a full-page wide table
\centering
\caption{Evaluation criteria for CRs tailored to CI contexts. While emphasizing CI‐specific needs such as ICS/OT fidelity and operational safety, the indicators and rubric anchors can be adapted for broader CR applications.}
\label{tab:criteria}
\setlength{\tabcolsep}{4pt}%
\renewcommand{\arraystretch}{1.05}%
\setlist[itemize]{noitemsep,topsep=0pt,leftmargin=*} % tighter lists inside cells
\resizebox{0.9\textwidth}{!}{%  <-- full two-column width
\begin{tabular}{>{\raggedright\arraybackslash}p{3cm} 
                >{\raggedright\arraybackslash}p{3.5cm} 
                >{\raggedright\arraybackslash}p{4cm} 
                >{\raggedright\arraybackslash}p{4cm} 
                >{\raggedright\arraybackslash}p{4.6cm}}
\toprule
\textbf{Criterion} & \textbf{Description} & \textbf{CI Applicability} & \textbf{Indicators} & \textbf{Rubric Anchors (1/3/5)} \\
\midrule

\emph{C1. Realism \& Fidelity} & Accurate replication of IT/OT stacks, protocols, timing, physics, and artifacts & Critical for CI skill transfer; reproduce ICS/SCADA, protocol semantics, and process effects. & Supported ICS protocols/devices; timing accuracy; HIL/physics simulators; realistic logs/traffic &
\begin{itemize}
  \item[1] IT-only, no ICS
  \item[3] Some ICS protocols/basic simulation
  \item[5] Rich ICS environment with physics/HIL and authentic artifacts
\end{itemize} \\
\addlinespace

\emph{C2. Security \& Isolation} & Safe containment of malware, tenant isolation, secure access, and auditability & CI scenarios may use sensitive data or real malware; leakage unacceptable & Network segmentation; virtualization isolation; RBAC/ABAC+MFA; separate red/blue spaces; logging/reset procedures &
\begin{itemize}
  \item[1] Weak isolation/shared credentials
  \item[3] Per-tenant environment + basic authentication
  \item[5] Hardened containment, full audit, strict operation policies
\end{itemize} \\
\addlinespace

\emph{C3. Scalability} & Capacity for users/nodes, stability under load, dynamic resource scaling & Sector drills may need hundreds of participants, long simulations & Max concurrent users/nodes; clustering/cloud burst; auto-scaling; long-duration stability; QoS &
\begin{itemize}
  \item[1] Small ($\leq$10 users), unstable
  \item[3] Medium ($\sim$50 users), manual scaling
  \item[5] Large ($\geq$200 users), smooth elastic scaling
\end{itemize} \\
\addlinespace

\emph{C4. Flexibility \& Extensibility} & Ease of scenario creation and integration of new assets/tools & CI sectors vary; adapt quickly to new devices, threats, and tools & SDK/APIs; import custom VM/container images; vendor-neutral libraries; external testbed interfacing &
\begin{itemize}
  \item[1] Closed, fixed templates
  \item[3] Configurable, limited plugins
  \item[5] Open APIs, rich libraries, hybrid integrations
\end{itemize} \\
\addlinespace

\emph{C5. Maintainability} & Effort to patch, update, and manage content & CI systems evolve slowly, threats fast, the CR must be easily updatable. & Patch/upgrade frequency; automated pipelines; infrastructure-as-code; version control; OSS vs proprietary &
\begin{itemize}
  \item[1] Manual, brittle updates
  \item[3] Semi-automated, regular updates
  \item[5] Full CI/CD, versioned artifacts, rollback
\end{itemize} \\
\addlinespace

\emph{C6. Usability} & Learnability and efficiency for instructors/trainees & CI trainees may be OT engineers; must be intuitive and operator-friendly. & GUI scenario builder; clear dashboards/HMIs; instructor controls; one-click resets; documentation &
\begin{itemize}
  \item[1] Steep learning curve, expert-only
  \item[3] Average usability
  \item[5] Highly intuitive, efficient management
\end{itemize} \\
\addlinespace

\emph{C7. Accessibility} & Practical access models (remote, thin client, federated) & CI exercises often cross organization/site boundaries; remote secure access key. & Web-based access; VPN/tunneling; low-bandwidth tolerance; SSO/federated auth; multi-tenancy &
\begin{itemize}
  \item[1] On-site only
  \item[3] Remote but requires a special client
  \item[5] Browser/zero-client, robust federated remote access
\end{itemize} \\
\addlinespace

\emph{C8. Training Effectiveness \& Measurement} & Telemetry, scoring, analytics, AAR, skills mapping & CI requires evidence of improved readiness and skill gap analysis & Built-in scoring engine; user action logging; timeline/AAR tools; skills framework mapping; reporting &
\begin{itemize}
  \item[1] Minimal/manual feedback
  \item[3] Basic scoring + replay
  \item[5] Rich analytics, dashboards, automated skill tracking
\end{itemize} \\
\addlinespace

\emph{C9. Cost \& Resource Efficiency} & Total cost vs. training value; efficiency of resource use & CI organizations face budget/resource limits; efficient scaling is key & License model; hardware/cloud cost; staff effort; use of virtual vs physical assets; node density per host &
\begin{itemize}
  \item[1] Very high cost, inefficient
  \item[3] Moderate cost, average efficiency
  \item[5] Excellent cost efficiency, flexible pricing
\end{itemize} \\
\addlinespace

\emph{C10. Vendor Support \& Ecosystem} & Support, documentation, training, updates, community & Ongoing vendor/community engagement ensures resilience and sector relevance & SLA; documentation; training/certs; update cadence; user community; sector-specific scenarios &
\begin{itemize}
  \item[1] Minimal support/docs, no ecosystem
  \item[3] Standard support + some scenarios
  \item[5] Robust SLAs, rich libraries, active CI community
\end{itemize} \\
\bottomrule
\end{tabular}}%  <-- closes \resizebox
\end{table*}

Each criterion of table~\ref{tab:criteria} can be evaluated using observable indicators and a 1–5 scoring rubric with descriptive anchors, outlined in the fourth and fifth columns of the table, respectively. Note that, for transparency, during criteria definition, evaluators should publish the detailed rubrics for each criterion, along with evidence or rationale for the assigned score. Each raw score is typically normalized to a 0–1 scale (or 0–100\%) for aggregation. In section~\ref{SS:AHP}, these scores are combined using weights, determined by AHP, to calculate an overall CR score. It is also recommended to conduct a sensitivity analysis on the weights, i.e., examining how changes in criteria importance affect the overall result, to ensure the evaluation is robust and transparent to the extent that each criterion influences the outcome.

\subsection{AHP for Criteria Weighting}
\label{SS:AHP}

To determine the relative importance of each evaluation criterion documented in table~\ref{tab:criteria}, we employ the AHP. It is a structured MCDA method that is well-suited for the needs of our study, i.e., CR evaluation. AHP was selected over alternative MCDA methods, such as TOPSIS~\cite{topsis}, ELECTRE~\cite{electre}, or simple weighted scoring~\cite{Kampourakis2025}, as it aligns with the study’s objectives. Particularly, it allows evaluators to express relative preferences through pairwise comparisons rather than absolute scores, including an internal consistency check to ensure logical coherence in expert judgments. 

In this way, these properties render it suitable for contexts like CR evaluation, where both qualitative and quantitative factors are combined in a transparent and explainable manner. Moreover, AHP’s widespread use in technology assessment, as mentioned in section~\ref{sec:Eval:apps} enhances comparability with existing studies. In the framework of figure~\ref{fig:framework}, evaluating the fitness of a CR for CI is the overall goal at the top of the hierarchy. Note that one could expand with sub-criteria beneath each main criterion and perform the AHP recursively per hierarchy layer, but we opt to treat each of the ten as a standalone factor for simplicity purposes. 

To address the challenge of determining weights through expert decision-making, we acknowledge a key assumption of our approach: instead of gathering direct input from a large pool of experts, which is both time-consuming and effort-intensive, we simulate the process computationally, with an LLM agent. Specifically, we leverage ChatGPT 5 \textit{deep research} feature's advanced analytical capability to simulate a panel of multidisciplinary experts. To ensure credibility and reproducibility, we provided the agent with the following detailed input prompt to guide the simulated expert panel's reasoning:

\begin{mdframed}[
    backgroundcolor=gray!5,
    linecolor=black!40,
    linewidth=0.6pt,
    roundcorner=6pt,
    innertopmargin=8pt,
    innerbottommargin=8pt,
    innerleftmargin=10pt,
    innerrightmargin=10pt,
    skipabove=10pt,
    skipbelow=10pt
]
\textbf{Simulated Expert Weighting Prompt.}  
You are to simulate a multidisciplinary panel of experts responsible for determining the relative importance of ten evaluation criteria for assessing CRs in Critical Infrastructure (CI) contexts. The simulated panel represents four complementary roles: (1) CR architect, (2) ICS/SCADA security engineer, (3) CI training coordinator, and (4) OT operations manager.  

\noindent The objective is to assess the relative importance of each criterion in evaluating a CR’s suitability for CI-focused training and assessment. The ten criteria are [\textit{Table~\ref{tab:criteria} is provided}].  

\noindent For every possible pair of criteria, reason step-by-step to determine which is more important from a CI perspective and to what degree, using Saaty’s 1–9 fundamental scale of relative importance. The reasoning should explicitly consider how each criterion influences key CI dimensions such as realism, safety, scalability, maintainability, and overall training effectiveness.  

\noindent Upon completion of the pairwise comparison matrix, verify the internal consistency of the judgments (Consistency Ratio, CoR~$<$~0.10). Finally, compute and report the normalized weight vector representing the relative importance of the ten criteria.
\end{mdframed}

Subsequently, the AHP weighting process initiates with pairwise comparisons among the ten criteria, guided by the simulated expert reasoning. Specifically, for each pair, the agent assesses their relative importance in the context of CI-oriented CR evaluation, deciding which criterion is more important and to what extent. To do that, Saaty’s 1–9 fundamental scale~\cite{saaty2001deriving} of relative importance is employed, where a value of 1 denotes nominal importance, 3 indicates moderate importance, 5 represents strong importance, 7 signifies very strong importance, and 9 reflects extreme importance of one criterion over the other. Intermediate values (2, 4, 6, 8) allow for finer distinctions, while reciprocal values express inverse judgments. These comparisons collectively form a positive reciprocal matrix that serves as the basis for deriving the priority weights.

All pairwise judgments populate an $n \times n$ comparison matrix $M$; in this case, $n$=10 criteria. Each entry $a_{ij}$ in the matrix represents the importance of the criterion $i$ relative to the criterion $j$. By definition, $a_{ji} = 1/a_{ij}$ for consistency, and all diagonal entries $a_{ii}=1$. Matrix~\ref{eq:ahpmatrix} shows the general form of such a matrix for $n$ criteria:

\begin{equation}
\label{eq:ahpmatrix}
\scriptsize
M = \begin{pmatrix}
1 & a_{12} & a_{13} & \cdots & a_{1n} \\
1/a_{12} & 1 & a_{23} & \cdots & a_{2n} \\
1/a_{13} & 1/a_{23} & 1 & \cdots & a_{3n} \\
\vdots & \vdots & \vdots & \ddots & \vdots \\
1/a_{1n} & 1/a_{2n} & 1/a_{3n} & \cdots & 1 \\
\end{pmatrix}
\end{equation}

Once the pairwise comparison matrix is constructed, AHP provides a systematic approach to derive a priority weight for each criterion. In essence, we seek a weight vector 
\(\mathbf{w} = (w_1, w_2, \dots, w_n)^\top\) 
that best represents the relative importance implied by the criteria comparisons. This can be obtained by solving the following eigenvalue problem:

\begin{equation}
M \mathbf{w} = \lambda_{\max} \mathbf{w}
\end{equation}

\noindent
where \(M\) is the \(n \times n\) pairwise comparison matrix, \(\lambda_{\max}\) is its principal eigenvalue, and \(\mathbf{w}\) is the corresponding eigenvector that expresses the relative priorities of the criteria. The resulting weight vector is then normalized so that the weights sum to one:

\begin{equation}
w_i = \frac{w_i}{\sum_{j=1}^{n} w_j}, \quad \text{for } i = 1, 2, \dots, n
\end{equation}

\noindent
ensuring that
\begin{equation}
\sum_{i=1}^{n} w_i = 1
\end{equation}

\noindent A key feature of the AHP is the logical consistency assessment of the experts’ judgments. For instance, if \(A > B\) and \(B > C\), it is logically expected \(A > C\), yet human judgments may not always satisfy this transitivity. To quantify the level of consistency, AHP computes a \textit{Consistency Index} (CoI) and a \textit{Consistency Ratio} (CoR). The CoI is derived from the maximum eigenvalue \(\lambda_{\max}\) of the comparison matrix \(M\) as follows:

\begin{equation}
CoI = \frac{\lambda_{\max} - n}{n - 1}
\end{equation}

\noindent
where \(n\) is the number of criteria. The CoI is then compared against a \textit{Random Index} (RoI), which represents the average CoI obtained from a large sample of randomly generated reciprocal matrices of order \(n\). CoR is calculated as:

\begin{equation}
CoR = \frac{CoI}{RoI}
\end{equation}

\noindent
As a rule of thumb, a CoR value of 0.10 or less (\(\leq 10\%\)) is considered acceptable, indicating that the judgments are reasonably consistent. If \(CoR > 0.10\), the judgments are deemed inconsistent, and the decision-makers revisit and revise the pairwise comparisons that contribute to the identified inconsistencies. Thus, in our framework, after the simulated expert panel produces the pairwise comparison matrix, we compute the CoR to validate the coherence of the derived weights. If the CoR exceeds 0.10, the agent flags the result, prompting a review and refinement of the inconsistent judgments. Note that the 0.10 threshold, as recommended by Saaty~\cite{SAATY1987161}, reflects an order-of-magnitude separation between the priority of consistency and the tolerable measurement error, preserving sensible accuracy while allowing minor inconsistency, allowing to capture evolving knowledge and judgment variability.

Once acceptable consistency is achieved, the final set of criteria weights \(\{w_i\}\) is established. For instance, a balanced panel of CI cybersecurity experts might assign the highest weight to \textit{Realism \& Fidelity}, as realistic ICS conditions are essential for meaningful training, a similarly high weight to \textit{Security \& Isolation}, given the paramount importance of safety, moderate weights to \textit{Training Effectiveness} and \textit{Flexibility}, and comparatively lower weights to \textit{Cost} or \textit{Vendor Support}, especially when capability is prioritized over price. These weights, however, are not fixed. They should be adapted to the specific priorities and mission objectives of the evaluators. For example, a utility operator may emphasize ICS realism, whereas a financial-sector entity might prioritize analytics and training effectiveness. The result of the AHP weighting process is a set of weights $w_1, ..., w_{10}$ summing to 1, corresponding to the ten criteria of table~\ref{tab:criteria}, respectively.

\noindent\textbf{Explainability of the Evaluation Process:} A key feature of the proposed framework is its explainability. Unlike black-box scoring or heuristic weighting methods, the AHP provides an explicit trace of how each criterion influences the final evaluation. Each pairwise comparison, whether generated by human or simulated experts (LLM agent), is documented and interpretable. The LLM-based simulated expert panel further contributes to explainability by providing natural-language rationales for each judgment, allowing evaluators to inspect and verify the reasoning that led to specific weights. Consequently, both the weighting and scoring phases of the framework are transparent and reproducible, supporting accountable decision-making in CR evaluation.

\subsection{Limitations}

Despite its merits, the proposed approach has certain methodological constraints and assumptions that should be acknowledged.

\noindent \textbf{L1:} Obviously, our LLM-driven approach has an inherent limitation; the model may generate plausible but inaccurate responses. Its feedback depends on how the questions are framed and lacks the contextual judgment and domain-specific insights that qualified experts provide. Additionally, ChatGPT’s knowledge is limited to its training data and may not reflect the most recent developments. In other words, we acknowledge that this LLM-based simulation cannot fully replace the judgment of human experts and experiential details; however, it provides a rational, transparent, and repeatable estimate of the relative importance of the criterion, serving as a practical proxy in early-stage framework validation.

\noindent \textbf{L2:} Another limitation regards the selected expert roles, i.e., CR architect, ICS/SCADA security engineer, CI training coordinator, and OT operations manager. We opt for these roles to capture the core technical and operational aspects relevant to CI-oriented CRs. For example, including other expert types (e.g., policy analysts or compliance auditors) would likely shift the resulting weights toward regulatory compliance, security assurance, or cost factors. Such variability is inherent to AHP, as it reflects the priorities of the participating decision-makers. Altogether, the current configuration serves as a technical baseline for CI-focused evaluations. Note, however, that the expert composition and corresponding weight distributions can be adapted, as needed.

\noindent \textbf{L3:} The framework has not yet been empirically validated through real-world expert panels or CR evaluations. Consequently, the assigned weights and criteria relationships should be regarded as illustrative rather than definitive until confirmed through broader expert engagement and case studies.

\section{Results}
\label{S:res}

As mentioned earlier, the simulation was performed using a ChatGPT 5 agent in deep research mode, acting as a panel of experts. It was informed by prominent guidelines and reports from organizations such as NIST~\cite{Nist:guide, SP-800-50, SP-800-82, nist:nice, nist:exp:lea}, ENISA~\cite{ECSO:check, european2020understanding}, U.S. DoD~\cite{DoD}, MITRE~\cite{mitre}, NATO~\cite{stevens2021cyber}, and various industry sources~\cite{cdt,IBM:cr,CR:hands-on,CR:TRYZUB} and academic sources~\cite{fi17060259, Kampourakis2025, app11041809, YAMIN2020}. This approach ensured that the pairwise comparisons reflect a consensus of current best practices.

\begin{equation}
\label{eq:pairwise-matrix}
M =
\scriptsize
\begin{bmatrix}
1     & 2     & 7     & 5     & 7     & 7     & 7     & 5     & 9     & 7     \\
1/2   & 1     & 7     & 5     & 7     & 7     & 7     & 3     & 9     & 7     \\
1/7   & 1/7   & 1     & 1/3   & 1/3   & 2     & 3     & 1/3   & 3     & 3     \\
1/5   & 1/5   & 3     & 1     & 2     & 3     & 3     & 1/3   & 5     & 3     \\
1/7   & 1/7   & 3     & 1/2   & 1     & 3     & 3     & 1/3   & 5     & 3     \\
1/7   & 1/7   & 1/2   & 1/3   & 1/3   & 1     & 3     & 1/5   & 3     & 3     \\
1/7   & 1/7   & 1/3   & 1/3   & 1/3   & 1/3   & 1     & 1/5   & 3     & 2     \\
1/5   & 1/3   & 3     & 3     & 3     & 5     & 5     & 1     & 7     & 5     \\
1/9   & 1/9   & 1/3   & 1/5   & 1/5   & 1/3   & 1/3   & 1/7   & 1     & 1/3   \\
1/7   & 1/7   & 1/3   & 1/3   & 1/3   & 1/3   & 1/2   & 1/5   & 3     & 1     \\
\end{bmatrix}
\end{equation}

The pairwise comparison matrix in Equation~\ref{eq:pairwise-matrix} represents how each evaluation criterion was judged relative to the others on Saaty’s scale of importance. Recall that each element \(M_{ij}\) indicates how much more important criterion \(i\) is compared to criterion \(j\), with reciprocal values (\(M_{ji} = 1/M_{ij}\)) ensuring consistency across comparisons. Overall, larger values in the upper triangle indicate criteria that dominate others in relative importance, while smaller reciprocals in the lower triangle indicate lower influence. Together, these relationships constitute the basis for computing the normalized priority weights shown in Table~\ref{tab:pairwise-matrix}.

Subsequently, the pairwise judgments were checked for consistency. The principal eigenvalue $\lambda_{\text{max}}$ of the matrix was calculated, and the CoI and CoR were determined using the standard AHP method. The consistency ratio was found to be well below 0.10, indicating an acceptable level of consistency in the comparisons.

\begin{minipage}{\columnwidth}
\centering
\captionof{table}{Pairwise comparison weights for the ten evaluation criteria, 
ordered by descending importance. $C_i$ denotes criterion $i$ as defined in Table~\ref{tab:criteria}.}
\label{tab:pairwise-matrix}
\scriptsize
\setlength{\tabcolsep}{4.5pt}\renewcommand{\arraystretch}{1.05}
\resizebox{0.8\columnwidth}{!}{%
\begin{tabular}{l c}
\toprule
\textbf{Criterion} & \textbf{Weight $w_i$} \\
\midrule
C1. Realism \& Fidelity                    & 0.317 \\
C2. Security \& Isolation                  & 0.254 \\
C8. Training Effectiveness \& Measurement  & 0.130 \\
C4. Flexibility \& Extensibility           & 0.079 \\
C5. Maintainability                        & 0.066 \\
C3. Scalability                            & 0.046 \\
C6. Usability                              & 0.039 \\
C7. Accessibility                          & 0.028 \\
C10. Vendor Support \& Ecosystem           & 0.025 \\
C9. Cost \& Resource Efficiency            & 0.016 \\
\bottomrule
\end{tabular}}%
\end{minipage}

Specifically, as depicted in table~\ref{tab:pairwise-matrix}, using the principal right-eigenvector method, we computed the normalized weight of each criterion. As expected, \textit{Realism \& Fidelity} (C1) concentrates the highest weight ($\approx$0.316), reflecting its paramount importance for CI-oriented CRs. \textit{Security \& Isolation} (C2) is the second-highest ($\approx$0.254), followed by \textit{Training Effectiveness \& Measurement} (C8) ($\approx$0.130). At the lower end, \textit{Usability} (C6) and \textit{Accessibility} (C7) are under 5\% each, \textit{Vendor Support \& Ecosystem} (C10) is $\approx$2.5\%, and \textit{Cost \& Resource Efficiency} (C9) has the smallest weight ($\approx$ 1.6\%). This aligns with industry guidance that cost should be considered only after critical capabilities and requirements are met. As pointed out earlier, for verification, we calculated the consistency index and ratio from the consensus pairwise matrix. 

The principal eigenvalue is calculated as $\lambda_{\max} \approx 10.92$, The CoI as $CoI = (10.92 - 10)/{9} \approx 0.102$, and using a random index \( RoI = 1.49 \) for \( n = 10 \), the CoR is $CoR = 0.102/1.49 \approx 0.069$. This value, \( CoR \approx 6.9\% \), is below the 10\% threshold for acceptable consistency, indicating a consistent set of judgments from the agent that simulated the experts.

\section{Framework Application}
\label{S:app}

In this section, we validate the applicability of the proposed evaluation framework by applying it to two representative CRs proposed for CI contexts. Specifically, we evaluate (i) \textit{PowerCyber}~\cite{Ashok2016}, a high-fidelity smart grid testbed that integrates real industrial hardware and SCADA systems, and (ii) \textit{ENIGMA}~\cite{SUHAIL2023}, an explainable digital twin–based CR for cyber–physical systems. Each range is assessed against the ten criteria defined in Table~\ref{tab:criteria}, weighted according to the weights in Table~\ref{tab:pairwise-matrix}.

\noindent\textit{\textbf{Use case I:}} \textit{PowerCyber} is a smart-grid CPS CR with actual industry hardware and real-time simulation. It integrates Siemens SCADA (SICAM PAS, Power TG), physical relays and PMUs, SCADA protocols (DNP3, IEC~61850, IEEE~C37.118), and HitL power simulators. Based on table~\ref{tab:criteria}, this matches the top realism rubric, so we assign C1=5. Furthermore, it utilizes virtualization and dedicated VM’s plus firewalls/VPN, giving per-tenant isolation with basic security, thus we assign C2=3. In terms of scalability, the authors note limited power simulator size and physical device count constraints, meaning that only a few concurrent users can use the range. This is well below a large commercial range; thus, we assign C3=2. Flexibility is also limited, as experiments are driven by predefined templates and GUI‐based device tools, so users cannot fully customize scenarios, assigning a value of C4=2. Maintainability is not discussed; no automated pipelines or updates are mentioned, assuming manual upkeep, thus, C5=1. Usability is moderate, as there is a comprehensive web UI for configuring attacks/defenses, but the underlying SCADA tools are complex and not designed for easy use. We conservatively score C6=3, denoting an average learning curve. Concerning accessibility, \textit{PowerCyber} is clearly built with a browser-based remote interface, getting a score of C7=5. Training support is minimal, as no built-in scoring or debrief engine is described; only raw logs and packet captures can be downloaded, matching the lowest rubric of C8=1. Regarding cost efficiency, the range relies on expensive physical hardware and simulators, so it falls under the C9=1 rubric. Vendor support is negligible; namely, \textit{PowerCyber} is an academic testbed with no commercial vendor or ecosystem, getting C10=1. Using the framework weights $w_i$ (Table~\ref{tab:pairwise-matrix}), the overall score is:

\[
\text{PowerCyber Score} = \sum_{i=1}^{10} w_i s_i \approx 3.28
\]

\noindent\textit{\textbf{Use case II:}} \textit{ENIGMA} describes a CR based on a digital twin of a vehicle’s CAN network. It does not include actual physical ICS devices, so its fidelity is moderate. It simulates realistic sensors, e.g., speedometer, indicator, coolant, etc., and CAN bus data in the Azure Digital Twin platform, but without actual hardware HitL. Based on table~\ref{tab:criteria}, rubric-wise, this is below \textit{rich HitL environment} but above \textit{IT-only}, so we assign C1=3. In terms of security and isolation, there is no mention of multi-user security or isolation, so C2=1, denoting no special isolation features. The platform uses Azure cloud services, which can scale to some extent, but it’s not designed for many concurrent trainees; we conservatively give C3=3. On the other hand, \textit{ENIGMA} is fairly flexible, as users can define custom DT models via the Digital Twin Definition Language, with open-source code, so we score C4=4. Maintainability is not discussed; we assume minimal automation, so we assign C5=2. Regarding usability, the gamified interface and XAI feedback through SHAP values help analysts, but a formal usability study is absent, so we give C6=3. Accessibility is considered high as \textit{ENIGMA} runs as a cloud service, effectively browser/web accessible, getting a score of C7=5. Training support is designed in a CTF-style training game with analytics, matching high rubric anchors, and getting a value of C8=4. Concerning cost efficiency, \textit{ENIGMA} uses cloud PaaS, namely, Azure Digital Twins, which incur some costs, depending on the resource consumption, so we assign C9=2. Vendor support/ecosystem is limited, as the prototype is open-source on GitHub, but it has no commercial backing, getting a score of C10=2. Using the framework weights $w_i$, as presented in Table~\ref{tab:pairwise-matrix}, ENIGMA’s overall score is:

\[
\text{ENIGMA Score} = \sum_{i=1}^{10} w_i s_i \approx 2.65
\]

\begin{figure}[t]
\centering
\includegraphics[width=0.8\linewidth]{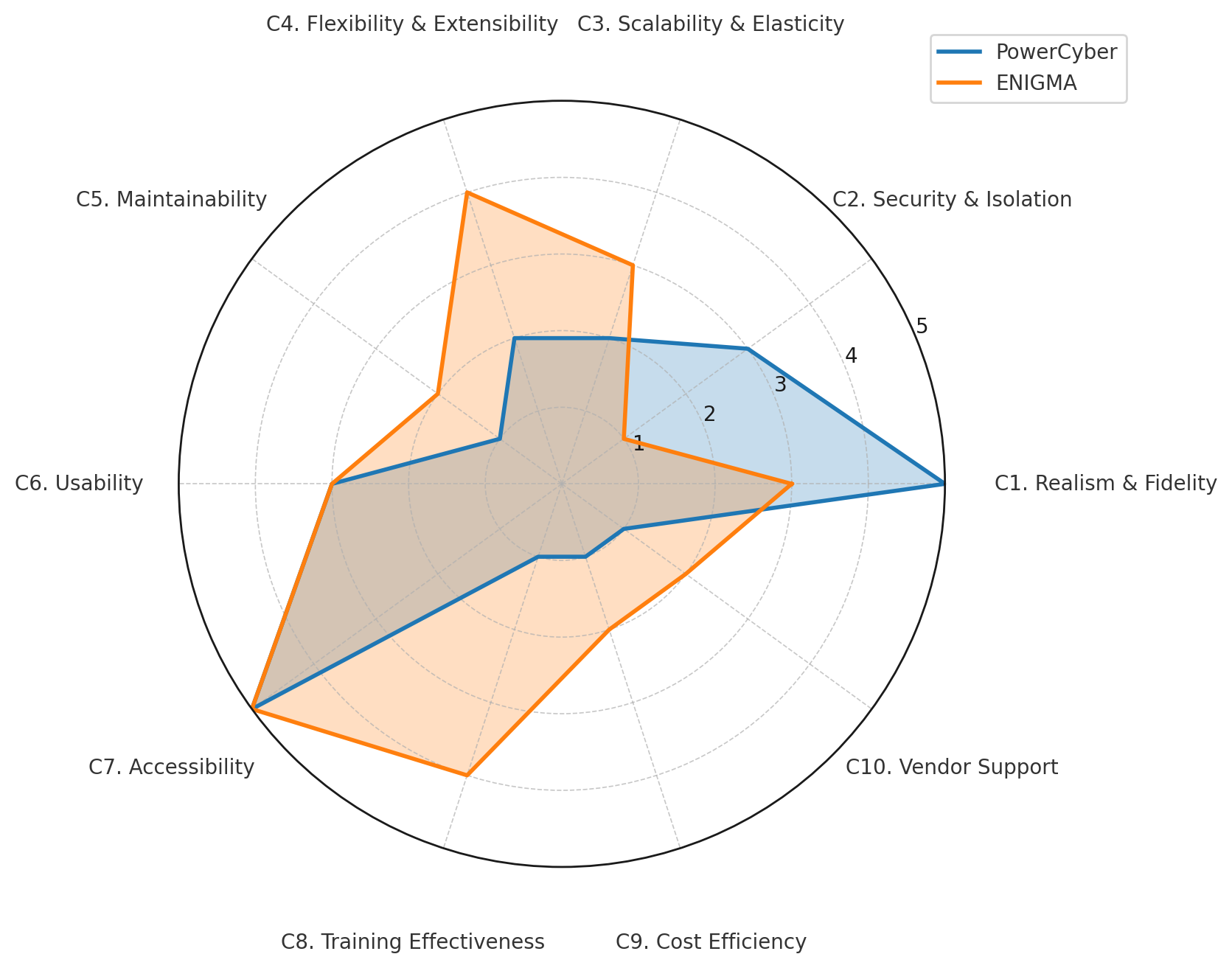}
\caption{Spider chart of the 10 criteria scores (1–5) for \textit{PowerCyber} and \textit{ENIGMA}.}
\label{fig:spider}
\end{figure}

For comparison, \textit{PowerCyber} excels in realism and accessibility against \textit{ENIGMA}, but scores deficiently in cost, maintainability, and training support. Contrarily, \textit{ENIGMA} trades off physical fidelity for flexibility and advanced training features. Overall, \textit{PowerCyber} achieves a higher weighted score ($\approx$3.28) than \textit{ENIGMA} ($\approx$2.65), reflecting \textit{PowerCyber’s} fidelity and production-grade realism at the expense of other factors. The individual criterion scores for both platforms are visualized in figure~\ref{fig:spider}.

\section{Conclusions}
\label{sec:conclusion}

CRs are gaining traction as indispensable platforms for cybersecurity education, training, and experimentation, offering controllable environments that can simulate or emulate complex IT and OT infrastructures. However, there is a scarcity of standardized methodologies for evaluation regarding the design, development, and refinement of CRs. In this direction, we proposed a structured evaluation framework tailored to CI-oriented CRs, integrating an MCDA approach with the AHP to systematically weight and aggregate diverse evaluation factors. A notable feature of the framework is the incorporation of simulated expert reasoning through an LLM agent, reducing the dependence on time-intensive human elicitation. The key contributions include: (1) a defined set of CI-specific evaluation criteria, (2) the application of AHP supported by LLM-based simulated expertise to ensure structured and explainable weighting, and (3) a practical scoring and aggregation mechanism yielding comparable results. The framework complements established guidelines by adding quantitative rigor, enabling organizations to assess not only \textbf{what} features are present but also \textbf{how well} they serve their mission. Future work should address current limitations by validating the framework with human expert panels and comparing outcomes across different LLMs to assess consistency and potential bias. Future work could also integrate reinforcement learning with expert feedback to refine the LLM’s decision process.

\section{Acknowledgments}

This work is supported by the Research Council of Norway through the SFI Norwegian Centre for Cybersecurity in Critical Sectors (NORCICS), project no. 310105 and by the European Union through the Horizon 2020 project PERSEUS (Grant No. 101034240). It has also received support from the European Union’s HORIZON Research and Innovation Programme under the ENFIELD project (European Lighthouse to Manifest Trustworthy and Green AI, Grant No. 101120657) and the European Regional Development Fund of the European Union under the CYBERUNITY project (Grant No. 101128024).

\bibliographystyle{ACM-Reference-Format}
\bibliography{sample-base}

\appendix

\end{document}